\documentclass[a4paper,10pt]{article}

\usepackage[latin1]{inputenc}
\usepackage[frenchb,english]{babel}
\usepackage[dvips]{graphicx}
\usepackage{these}

\begin{document}
\sloppy

\noindent {\bf Cybergeo : Revue européenne de géographie, N°295, 06 décembre 2004}

\begin{center}
	{\large \bf Modélisation prospective de l'occupation du sol. Le cas d'une montagne méditerrannéenne.}
\end{center}
\begin{center}
	(Prospective modelling of georeferenced data by crossed GIS and statistic approaches applied to land cover in Mediterranean mountain areas)
\end{center}
\hspace*{1.5 cm}
\begin{center}
	Martin Paegelow\footnote[1]{\texttt{paegelow@univ-tlse2.fr} - GEODE, UMR 5602 CNRS, 5 allées Antonio Machado, 31058 Toulouse cedex 9}, Nathalie Villa\footnote[2]{GRIMM, équipe d'accueil 3686, 5 allées Antonio Machado, 31058 Toulouse cedex 9} Laurence Cornez\footnote[3]{Stagiaire au GEODE, UMR 5602 CNRS}, Frédéric Ferraty\footnotemark[2], Louis Ferré\footnotemark[2], Pascal Sarda\footnotemark[2]
\end{center}

\selectlanguage{frenchb}
\begin{petitresume}
	Les auteurs mettent en \oe uvre trois méthodes de modélisation prospective appliquées à des données géoréférencées haute résolution portant sur l'occupation du sol en milieu montagnard méditerranéen : approche SIG, modèle linéaire généralisée et réseaux neuronaux. Une validation des modèles est entreprise par la prédiction de l'occupation du sol à la dernière date connue. Les résultats obtenus sont, dans le contexte de la dynamique spatio-temporelle de systèmes ouverts encourageants et comparables. Les scores de prédiction correcte se situent autour de 73 \%.  L'analyse des résultats porte notamment sur la localisation géographique, les types d'occupation du sol concernés et les écarts à la réalité des résidus. Un croisement des trois modèles souligne le degré élevé de convergence et une relative similitude des résultats issus des deux approches statistiques comparée au modèle SIG supervisé. Des travaux en cours concernent la mise en \oe uvre des modèles sur d'autres sites et le repérage des points forts respectifs afin de développer un modèle intégré.    
\motscles{modélisation, modèle linéaire généralisé, prévision, réseaux de neurones,
SIG}
\end{petitresume}

{\selectlanguage{english}
\begin{littleabstract}
	The authors apply three methods of prospective modelling to high resolution georeferenced land cover data in a Mediterranean mountain area: GIS approach, non linear parametric model and neuronal network. Land cover prediction to the latest known date is used to validate the models. In the frame of spatial-temporal dynamics in open systems results are encouraging and comparable. Correct prediction scores are about 73 \%. The results analysis focuses on geographic location, land cover categories and parametric distance to reality of the residues. Crossing the three models show the high degree of convergence and a relative similitude of the results obtained by the two statistic approaches compared to the GIS supervised model. Steps under work are the application of the models to other test areas and the identification of respective advantages to develop an integrated model.
\end{littleabstract}}

\selectlanguage{francais}

\selectlanguage{frenchb}

\section{Problématique et objectifs}

L'objet de notre recherche est la modélisation de dynamiques environnementales dans le cadre de systèmes complexes et ouverts. Dans ce cadre, la variable étudiée - ainsi que les variables d'environnement, susceptibles d'expliquer son évolution dans l'espace et dans le temps - contient une part d'incertitude ou d'aléa ce qui exclut, de fait, une approche déterministe. Ainsi, nous utilisons une approche stochastique (probabiliste) tenant compte de la dépendance dans le temps (effet mémoire) et dans l'espace : les outils probabilistes utilisés sont notamment la distribution multinomiale et l'analyse de Markov. En outre, notre approche fait également appel à la logique floue et à un automate cellulaire. L'ensemble de ces méthodes est mis en \oe uvre dans trois modèles à but prévisionnel différents afin de comparer leurs performances respectives. Plus précisément, il s'agit de comparer un modèle géomatique combiné de simulation prospective dont la mise en \oe uvre est possible en utilisant les fonctions de logiciels SIG disponibles sur le marché à deux modèles statistiques dont la mise en \oe uvre, plus longue, est extérieure au SIG. En contrepartie, l'intérêt des deux approches statistiques réside dans leur caractère automatique tandis que le modèle SIG nécessite une analyse thématique experte.

Notre choix en matière de modélisation statistique a porté sur deux approches classiques, l'une basée sur le maximum de vraisemblance (modèle linéaire généralisé) et l'autre utilisant un réseau de neurones. Ces deux méthodes sont proches du point de vue de la modélisation et différent essentiellement en ce qui concerne les algorithmes de mise en \oe uvre.

Un des défis actuels de la recherche en géomatique est celui de la modélisation prospective de données géographiques à haute résolution. Des méthodes géostatistiques éprouvées pour l'interpolation et l'extrapolation spatiales existent depuis plusieurs décennies et sont implémentées dans nombre de logiciels géomatiques commercialisés. Par contre, des outils de modélisation temporelle et d'aide à la décision ne furent implémentés dans les SIG que récemment et doivent être considérés plutôt comme des algorithmes expérimentaux intéressants que de techniques opérationnelles.

Depuis les années 1990 la demande sociale en outils d'aide à la décision et de modélisation capables d'assister différentes tâches de gestion environnementale (notamment la prévention de risques) et d'aménagement des territoires s'est fortement accrue.

Cet article illustre les premiers résultats d'une étude comparative de trois méthodes de modélisation prospective appliquée à l'occupation du sol dans des anthroposystèmes montagnards de l'Europe du sud. Nous considérons l'occupation du sol comme un indicateur pertinent, disponible à haute résolution, d'une combinaison d'activités humaines que les sociétés déploient dans l'espace - et auxquelles l'occupation du sol réagit avec une certaine inertie - et de facteurs naturels.
Les montagnes méditerranéennes font l'objet d'une profonde restructuration socio-économique qui se manifeste, entre autres, dans de spectaculaires changements paysagers. Cette réorganisation commença dans les Garrotxes (Pyrénées Orientales, zone d'études) à la fin de la première moitié du XIX$^\textrm{ème}$ siècle par le déclin du système agropastoral traditionnel provoquant l'exode rural. 

Une base de données géoréférencées matérialise les connaissances des dynamiques passées et actuelles de l'occupation du sol ainsi que des facteurs d'environnement potentiellement explicatifs. Elle alimente trois méthodes de modélisation prospective : l'une, supervisée, met en \oe uvre des algorithmes implémentés dans des logiciels SIG; les deux autres approches - modèle linéaire généralisé et réseau neuronal - peuvent être qualifiées de non supervisées dans la mesure où les règles du comportement spatio-temporel sont automatiquement détectées par l'outil. L'objectif principal étant la mise en \oe uvre et l'optimisation de chacune des trois approches sur le même jeu de données. L'interprétation critique des résultats, notamment des résidus de la prédiction, permet de cerner avantages et inconvénients respectifs. A partir de cette analyse comparative, peuvent être envisagées la possibilité de construction d'un modèle optimisé intégrant les points forts de chacune des méthodes ainsi que les modalités de transposition et les limites de généralisation.

Afin de valider et d'optimiser les modèles ceux-ci sont appliqués, dans un premier temps,  à prédire l'occupation du sol à la dernière date connue avant de proposer des scénarii prospectifs. Cette étude est menée dans le cadre d'une coopération entre trois équipes de recherche travaillant sur deux sites\,\footnote{GEODE UMR 5602 CNRS, Groupe SMASH -EA 3686 GRIMM, UTM
 et Instituto de Désarrollo Régional - Universidad de Granada} : les Garrotxes dont nous présentons ici les premiers résultats et la Alta Alpujarra Granadina (Andalousie, Espagne - travaux en cours).

\section{Zone d'études et base de données}
\subsection{Les Garrotxes}

Les Garrotxes, situées dans le département des Pyrénées Orientales, à l'extrémité NO du Conflent forment un ensemble géographique constitué de cinq communes et d'une taille de 8 570 ha. Ce bassin versant présente une rive droite granitique, à modelé géomorphologique relativement lourd, où sont localisées la quasi-totalité des anciennes terrasses de cultures et des forêts de pins à crochet (Pinus uncinata) et de pins sylvestre (Pinus sylvestris) ; un espace à dynamique végétale très rapide. La rive gauche du Cabrils, cours d'eau collecteur ce jetant dans la Têt à Olette, est un large soulane orientée SO sur substrat schisteux avec un métamorphisme de contact dans les zones les plus basses et occupée par des landes majoritairement ligneuses (à base de Genista purgans et, dans une moindre mesure, de Calluna vulgaris), fortement embroussaillée aux altitudes les plus basses par des chênes verts (Quercus ilex). La particularité des Garrotxes est leur enclavement : le bassin versant, à l'écart des grandes routes, est délimité au nord par le massif du Madrès (2469 m), à l'ouest (Cami Ramader) et au sud (Puig de la Tossa, Serrat del Cortal) par des chaînes culminant entre 1600 et 2000 m d'altitude et à l'est par la crête (Lloumet) de la soulane  rejoignant le Madrès. La vallée du Cabrils présente une dégradation progressive du climat méditerranéen ; la remontée de l'influence méditerranéenne au c\oe ur des Pyrénées Orientales étant assurée par la vallée de la Têt modifiant ainsi la rudesse du climat montagnard.

\begin{figure}[h]
\begin{center}
\includegraphics[width=12 cm]{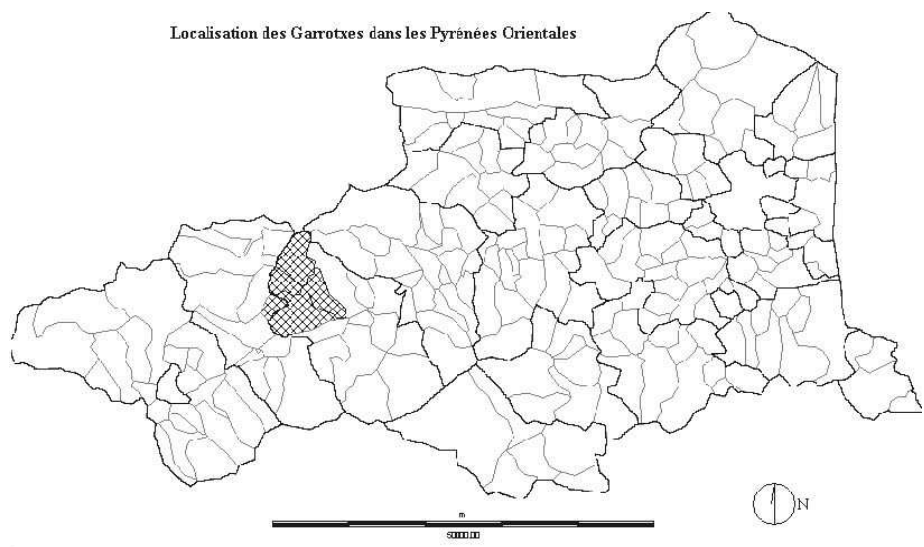}
\end{center}
\caption{\label{geo2_localisation} Localisation des Garrotxes à l'intérieur du département des Pyrénées Orientales}
\end{figure}

Autrefois un modèle d'organisation agropastorale traditionnelle, de nos jours l'agriculture a quasiment disparu tandis que l'activité pastorale, longtemps en déclin, donne des signes de renouveau suite à une profonde réorganisation entamée durant les années 1980 (\cite{metailie_paegelow_DPC2004}, \cite{paegelow_camacho_PACDRVMMM2004}). Le maximum démographique au début du XIX$^\textrm{ème}$ siècle se traduisait par une mise en valeur de toutes les ressources montagnardes traditionnelles mobilisables (agriculture, élevage, sylviculture). Ainsi en 1826 (cadastre napoléonien) un quart de la surface totale était cultivé. Le support quasi exclusif de l'agriculture a été les terrasses de cultures (feixtes). Le déclin démographique (de 1 832 habitants en 1830 à 90 en 1999) et la reconversion des terrasses de culture en pâturages, broussailles et forêts allaient de pair.

Parmi les agents externes considérés responsables du déclin de cette société locale à faible degré d'insertion dans l'économie nationale on peut citer, outre les processus d'industrialisation et de mise en valeur agricole de plaines au cours du XIX$^\textrm{ème}$  siècle, une variabilité interannuelle accrue des précipitations, observée au milieu du XIX$^\textrm{ème}$  siècle (\cite{tabeaud_pech_simon_DEHDM2003}), qui eût pu contribuer à la rupture d'un système poussé à bout par la pression anthropique sur le milieu. Deux évènements ponctuels - l'arrivée du chemin de fer à Olette (1911) et la Première Guerre Mondiale - ont accéléré l'exode rural. Ainsi il est probable que l'avenir proche se jouera en termes de gestion - ou de non gestion - pastorale se matérialisant par divers moyens de blocage, voire d'inversion, de l'embroussaillement et du reboisement spontané des espaces pastoraux (berger guidant le bétail, clôtures, écobuage). L'instauration de groupements pastoraux (GP) et d'associations foncières et pastorales (AFP) à partir des années 1980 a effectivement conduit à une reprise de l'activité pastorale avec un remplacement partiel du cheptel ovin par des bovins et des équins. Les signes de reconversion économique sont récents (années 1990) mais d'une portée limitée (ouverture d'un gîte d'étape à Sansa, tentatives de valorisation en tourisme vert) malgré la concrétisation prévue prochainement du Projet de Parc Naturel Régional des Pyrénées Catalanes.

\subsection{La base de données et l'évolution de l'occupation du sol}
\label{geo2_base donnees}
La base de données géoréférencées consiste en une série de cartes d'occupation du sol assorties de plans d'information représentant des facteurs environnementaux et sociaux. Les cartes existent - selon les traitements envisagés - soit en mode image (résolution du pixel d'environ 18 m), soit en mode objet. Ainsi les principaux traitements pour la modélisation font appel à une logique image (analyse spatiale) tandis que le mode objet offre une plus grande souplesse pour réaliser des requêtes attributaires.

Les cartes d'occupation du sol ont la même résolution spatiale mais ont des origines et des légendes variables. La première carte est basée sur le cadastre napoléonien (1826) - un support permettant de distinguer entre forêts, espaces pastoraux, prairies, champs agricoles et le bâti (villages). La première mission aérienne disponible (1942) rend possible le renseignement de la catégorie broussailles (landes très embroussaillées contenant de groupes d'arbres ou un nombre important d'arbres isolés) - maillon manquant en 1826 entre les formations arborescentes denses et arbustives (landes ligneuses). La carte de 1962 conserve la même légende tandis que l'échelle et la qualité des missions aériennes plus récentes (1980 et 1989), également panchromatiques, facilitent la distinction entre forêts de conifères et formations boisées de type feuillus. Il en va de même pour une meilleure discrimination des espaces pastoraux : landes ligneuses (notamment à base de Genista purgans) et landes à graminées. La carte d'occupation du sol la plus récente (2000) est basée sur des observations de terrain et est, par conséquent, nettement plus détaillée (20 catégories).

En raison de la nature des sources, la classification de l'occupation du sol, sous forme de trois nomenclatures emboîtées,  est surtout d'ordre physionomique.

\begin{figure}[h]
\begin{center}
\includegraphics[width=12 cm]{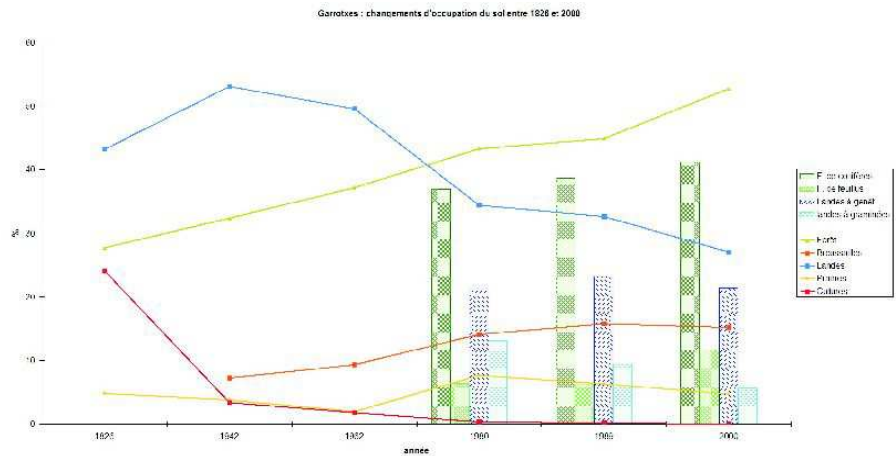}
\end{center}
\caption{\label{geo2_repartition} Changements de l'occupation du sol entre 1826 et 2000 (Garrotxes)}
\end{figure}

L'évolution de l'occupation du sol (cf. Figure \ref{geo2_repartition}) est classique. Les terres labourées délaissées ont été d'abord utilisées comme zones de pâture avant un embroussaillement menant souvent à la reconquête par la forêt. La base de données géoréférencées contient nombre de plans d'information ayant trait à l'occupation du sol :
\begin{itemize}
\item[$\bullet$] Plans issus du MNT : carte altitudinale, carte des pentes, carte d'occupation du sol ;
\item[$\bullet$] Plans d'accessibilité calculés à partir du réseau routier et l'emplacement des villages (habitat groupé) : indice d'accessibilité (analyse de distance-coût) selon la date ;
\item[$\bullet$] Plans relatifs à la gestion pastorale : unités pastorales (UP), associations foncières et pastorales (AFP), pression pastorale ;
\item[$\bullet$] Plans tenant compte du statut particulier de certains zones : forêts domaniales et zone militaire ;
\item[$\bullet$] Limites administratives ;
\item[$\bullet$] Réseau hydrographique.
\end{itemize}

Les données purement attributaires (recensement de la population, recensement général agricole, \ldots) sont, selon leur degré de confidentialité, connus, mais apportent peu de connaissances compte tenu de leur unité spatiale de rattachement (la commune) incompatible avec une analyse haute résolution.

\section{Méthodologie et mise en \oe uvre}
Il s'agit de construire trois modèles prédictifs de l'occupation du sol (variable qualitative) à haute résolution, de les calibrer par un test sur la dernière date connue en utilisant la même base de données afin de comparer leurs efficacités respectives. La première approche fait appel aux techniques disponibles dans les SIG et peut être qualifiée de supervisée dans la mesure où l'analyse du géographe conduit à l'implémentation des règles nécessaire au calcul des cartes de probabilité. Bien que les deux autres approches, statistiques, soient également supervisées, le rôle du mathématicien, peu familier avec la thématique, consiste plutôt à optimiser l'algorithme que d'y introduire les conclusions de sa propre analyse du comportement spatio-temporel du milieu considéré.
\subsection{Approche supervisée par SIG}
Le modèle que nous présentons se veut être simple à deux égards : simple dans le lever des données d'entrée (limitation à quelques données facilement disponibles, cf. paragraphe \ref{geo2_base donnees}) et simple dans sa mise en \oe uvre informatique (recours à des algorithmes implémentés dans un logiciel SIG commercialisé). Ce modèle géomatique combiné :

\begin{itemize}
\item[$\bullet$] Fait appel à la logique floue afin d'ajuster les données environnementales dans l'évaluation multicritère.
\item[$\bullet$] Est stochastique pour l'aspect prédictif de la simulation à événements discrets et états finis - chaînes de Markov avec mémoire (deux dates initiales).
\item[$\bullet$] Remédie aux limites de l'analyse markovienne en recourant à une évaluation multicritère optimisant l'affectation spatiale des probabilités markoviennes (prise en compte de la rugosité de l'espace) par constitution d'une base de connaissances et de règles d'inférence (variables d'environnement) relatives à la phase d'apprentissage (calibration) du modèle.
\item[$\bullet$] Utilise un automate cellulaire simple pour favoriser l'émergence de zones de probabilités d'états à extension surfacique réaliste.
\end{itemize}

Mise en \oe uvre sous le logiciel Idrisi 32, la modélisation se découpe en trois phases :

\begin{itemize}
\item[$\bullet$] La constitution de la base de connaissances de la dynamique spatio-temporelle de l'occupation du sol par évaluation multicritère (EMC) des variables d'environnement. Les variables d'environnement d'origine sont transformées, pour chacun des types d'occupation du sol, par traitements statistiques et par logique floue en plans de probabilité d'occurrence de chacune des catégories d'occupation du sol. Ces plans de probabilité résultants, se basant sur la période d'apprentissage (1980 à 1989) servent à l'allocation spatiale des probabilités de transition.
\item[$\bullet$] Le calcul des probabilités de transition par analyse de chaînes de Markov (ACM) entre les dates de la phase d'apprentissage et la date simulée (2000 - dernière date connue).
\item[$\bullet$] L'allocation spatiale des probabilités de transition markoviennes : cette dernière étape utilise les résultats catégoriels de l'EMC. Ceux-ci sont intégrés, par évaluation multiobjectif (EMO), en une seule carte de l'occupation du sol simulée laquelle est traitée par un automate cellulaire (AC)  basé sur un filtre de contiguïté spatiale.
\end{itemize}

La calibration du modèle est obtenue en modélisant l'état de l'occupation du sol à la dernière date connue (2000) sur la base de l'information provenant d'une période d'apprentissage englobant les deux dates précédentes (1980 et 1989). Bien que nous disposons d'une connaissance historique plus approfondie, il est évident que les conditions socio-économiques actuelles ne s'appliquent pas aux états de l'occupation du sol du XIX$^\textrm{ème}$ et du XX$^\textrm{ème}$ siècle jusque dans les années 1960.

\subsubsection{Construction d'une base de connaissances des dynamiques de l'occupation du sol}

La connaissance des dynamiques récentes est essentielle pour appréhender l'évolution future et sa modélisation. Nous entendons par connaissance des mesures statistiquement significatives du comportement spatial et temporel de l'occupation du sol en relation avec des critères environnementaux considérés explicatifs d'une partie de sa variabilité. Dans une évaluation multicritère (EMC), on distingue entre des critères binaires, les contraintes, et les critères ayant une aptitude variable dans l'espace, les facteurs. Les contraintes booléennes masquent certaines zones (occurrence possible ou non) ; ils peuvent s'appliquer à toutes formes d'occupation du sol (exclusion des espaces bâtis) ou être spécifique à certaines formations (limite altitudinale des conifères). Les facteurs traduisent une connaissance graduée pour l'objectif en question (une forme d'occupation du sol) - ils peuvent être pondérés et on peut définir leur degré de compensation.

Pour chaque catégorie d'occupation du sol la connaissance de son comportement spatio-temporel provient d'une analyse diachronique des dynamiques et de la friction géographique en comparant la répartition théorique (espace homogène) et la répartition réelle (niveaux de probabilité 99\% et 99.9\%). La rugosité géographique est exprimée par les facteurs d'environnement cartographiés (altitude, pente, exposition, accessibilité, proximité aux entités de même nature, statut de gestion particulière de certaines zones et probabilité de changement) et disponibles à la même résolution spatiale.

\begin{figure}[h]
\begin{center}
\includegraphics[width=6 cm]{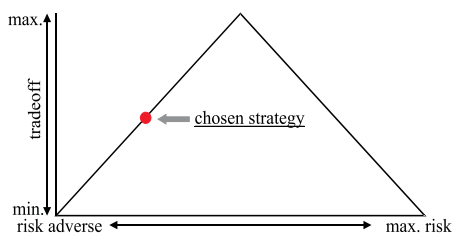}
\end{center}
\caption{\label{geo2_strategie} Espace de décision et approche EMC-OWA choisie}
\end{figure}

Les facteurs sont standardisés par recodage manuel ou par l'emploi de fonctions fuzzy et pondérés par l'emploi de la matrice de Saaty (\cite{saaty_JMP1977}) qui renvoie le vecteur propre de chaque facteur. L'approche de l'EMC développée (\cite{eastman_kyem_toledano_jin_EGST1993}) inclut des poids d'ordre (ordered weighted averaging - OWA) permettant le choix du niveau de risque et de compensation entre facteurs. Ces poids d'ordre classent les aptitudes par rang croissant et aboutissent à un classement spécifique à chaque pixel. Nous avons opté pour une approche conservatrice (peu de risques et niveau de compensation limité) comme l'exprime la Figure~\ref{geo2_strategie}.

\subsubsection{Calcul des probabilités de transition}

Le calcul prédictif de l'occupation du sol est opéré par une analyse des chaînes de Markov, un processus discret avec des pas temporels discrets et dont les valeurs à la date prédite dépendent des valeurs à des dates antérieures. La prédiction est exprimée par une estimation des probabilités de transition.
Le test consiste à prédire l'occupation du sol en 2000 (dernière date connue) sur la base de 1980 et de 1989. Le résultat se présente sous forme d'une matrice dans laquelle sont codées les probabilités de changement de chaque catégorie d'occupation du sol ainsi que le nombre de pixels affectées entre la dernière date d'apprentissage et la date projetée. La fonction calcule également une carte de probabilité conditionnelle pour chaque catégorie d'occupation du sol indiquant la probabilité markovienne par pixel de la modalité en question à la date projetée. Une intégration stochastique de l'ensemble des cartes par modalité en une seule est possible mais aboutit à un résultat relativement éloigné de la réalité (image bruitée) car cette procédure purement statistique ne tient pas compte ni des règles de connaissances établies par EMC, ni de la continuité spatiale.

\subsubsection{Allocation spatiale des probabilités prédites}

\begin{figure}[h]
\begin{center}
\includegraphics[width=10 cm]{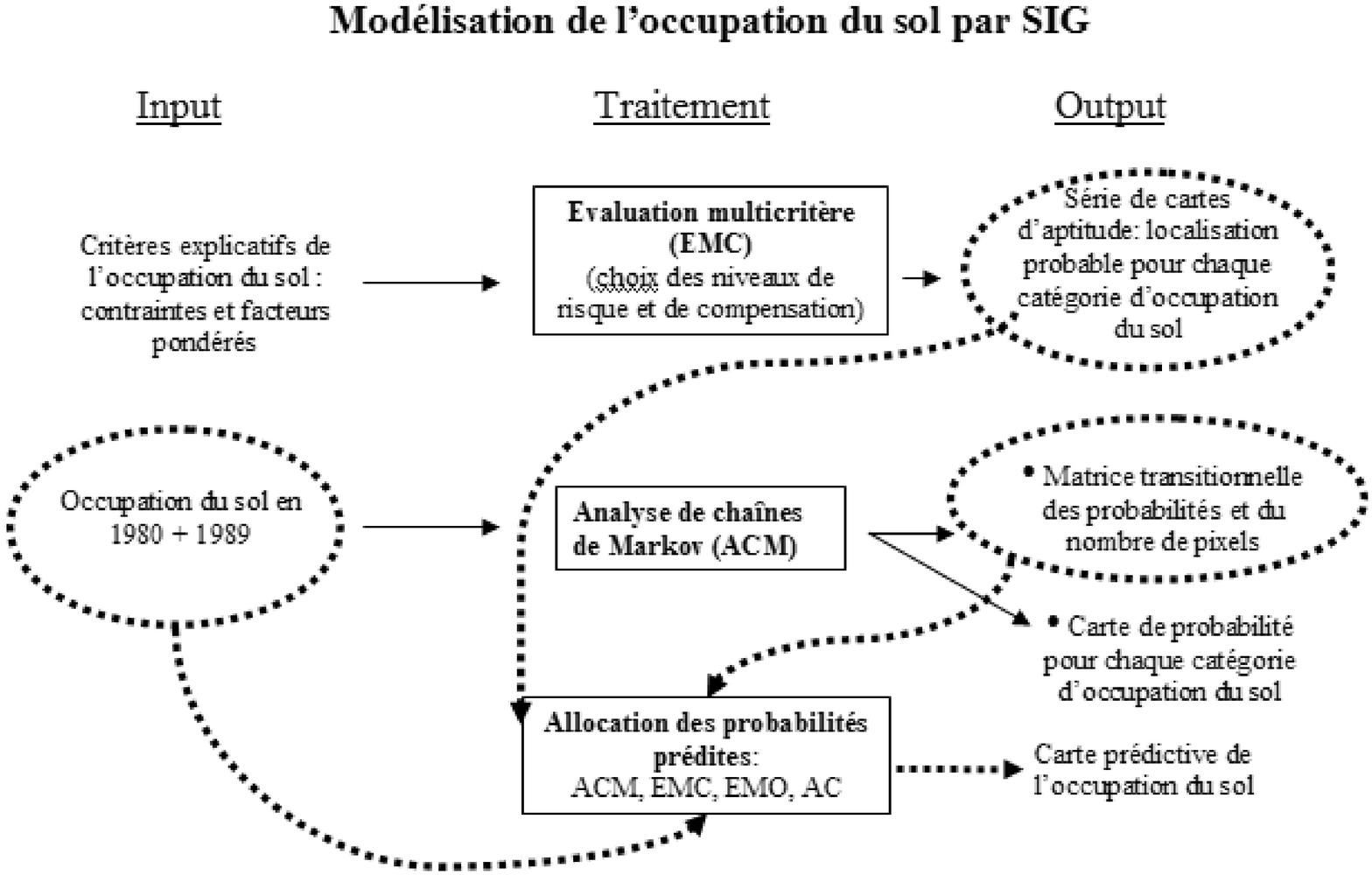}
\end{center}
\caption{\label{geo2_modelisationSIG} Modélisation de l'occupation du sol par SIG : aperçu des principales fonctions et de leur enchaînement}
\end{figure}

L'allocation spatiale des probabilités markoviennes intègre la connaissance sur la répartition probable de l'occupation du sol (EMC), une évaluation multiobjectif (EMO) tenant compte des objectifs concurrents (chaque modalité d'occupation du sol étant un objectif) et un automate cellulaire, basé sur un filtre de contiguïté spatiale. La fonction utilisée sous Idrisi est CA\_Markov dont l'algorithme est itératif afin de tenir compte des distances temporelles entre les deux dates d'apprentissage et la dernière date d'apprentissage et la date de projection. Elle donne en sortie une carte prospective de l'occupation du sol probable.
La Figure \ref{geo2_modelisationSIG} résume les principales étapes de modélisation par SIG (\cite{paegelow_ACG2003}, \cite{paegelow_camacho_menorMPPMSIG2004}).

\subsection{Approche par réseaux de neurones}

L'idée de l'utilisation de réseaux de neurones a été principalement motivée par leurs remarquables capacités d'adaptation et de souplesse face à un très grand nombre de problèmes, notamment lorsque ceux-ci présentent des aspects non linéaires ou lorsque les variables explicatives sont fortement corrélées ; ces deux aspects se retrouvent dans le travail de prédiction que nous cherchons à effectuer. Aussi, les réseaux de neurones ont récemment connu une grande popularité et ont très favorablement concurrencé les méthodes statistiques classiques. On les retrouve notamment dans la prédiction de séries chronologiques (cf. \cite{bishop_NNPR1995}, \cite{lai_wong_JASA2001} et \cite{parlitz_merkwirth_esann2000}). Le cadre dans lequel nous sommes amenés à travailler est encore plus étendu puisqu'il s'agit ici d'un processus spatio-temporel auquel s'ajoutent des variables explicatives comme nous le détaillerons plus loin.

\subsubsection{Réseaux de neurones multi-couches (perceptrons)}
\begin{figure}[h]
\begin{center}
\includegraphics[width=10 cm]{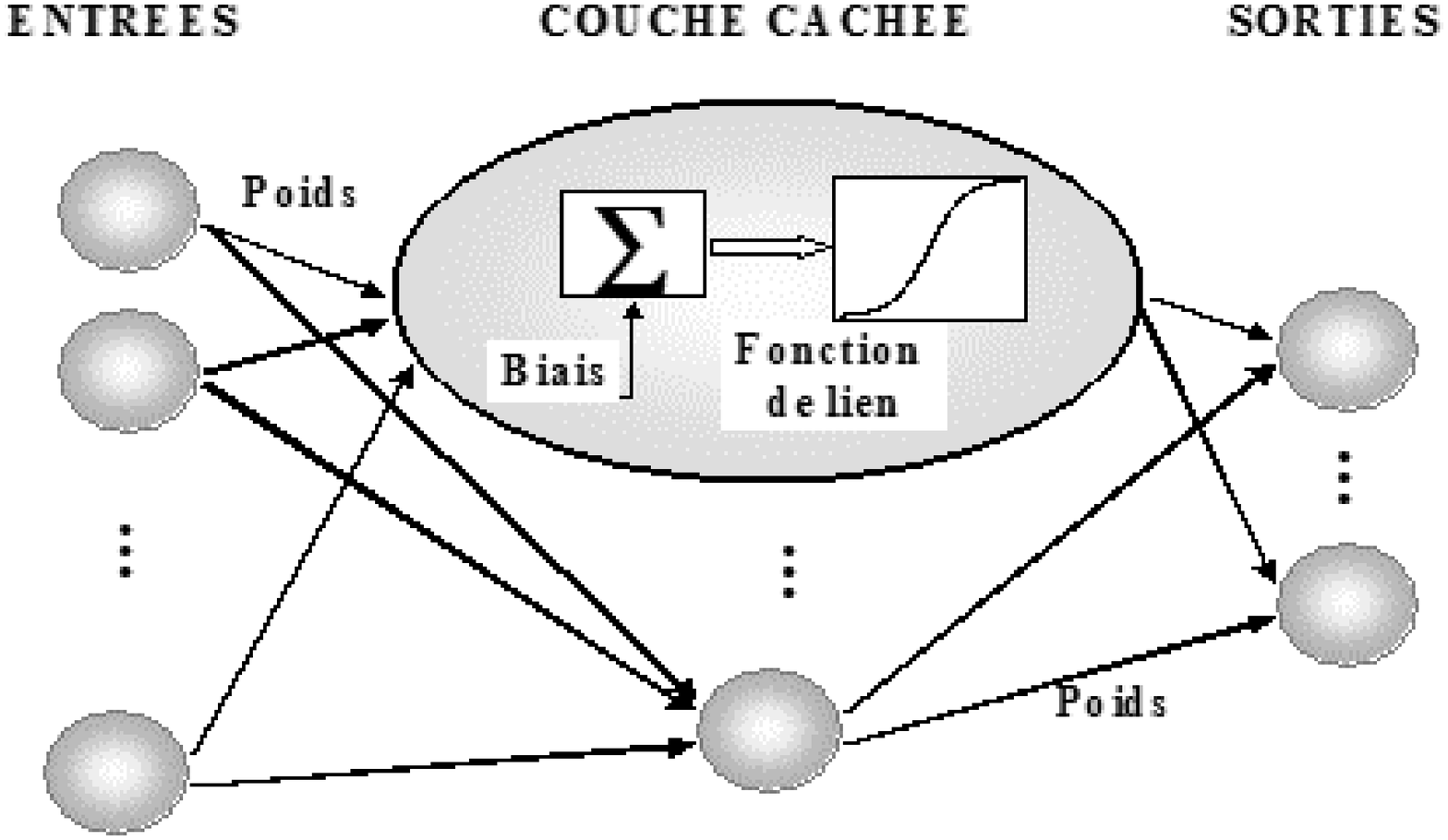}
\end{center}
\caption{\label{geo2_nn} Architecture d'un réseau à une couche cachée}
\end{figure}

Nous avons travaillé avec une classe particulière de réseaux de neurones, les réseaux multi-couches ou perceptrons. Ceux-ci ont été les premiers à connaître un essor important ; leur création est issue des premières tentatives de modélisation des principes de base régissant le fonctionnement du cerveau même si leur champ d'application s'est, depuis, considérablement élargi, notamment au traitement de données statistiques (pour plus de détails, consulter \cite{davaloe_naim1969} ).
Lorsque l'on parle de réseaux à couches, le nombre de couches est à définir par l'utilisateur mais doit comprendre au minimum une couche d'entrée et une couche de sortie ; les autres couches dont le nombre varie sont appelées couches cachées ; pour leurs remarquables propriétés d'approximation, nous avons choisi d'utiliser un réseau à une couche cachée dont l'architecture générale est celle de la Figure \ref{geo2_nn}.

Détaillons un peu le fonctionnement de ce réseau : en entrée, la valeur des neurones est celle des variables explicatives du modèle ; chacune de ces valeurs numériques est multipliée par un certain nombre de poids pour être, finalement, additionnée et transformée par une fonction de lien au niveau des neurones de la couche cachée. Enfin, les valeurs numériques des neurones de la couche cachée subissent à leur tour une multiplication par des poids et leur addition donne la valeur des neurones de sortie qui modélisent la variable expliquée. Les poids, généralement notés w, sont choisis lors d'une phase dite d'apprentissage sur un jeu de données test et minimisent l'erreur quadratique de ce jeu de données. Finalement, les réseaux de neurones à une couche cachée sont les fonctions de la forme :
$$
\Psi_w(x) = \sum_{i=1}^{q} w_i^{(2)} g\left(x^Tw_i^{(1)} + w_{i,0}^{(1)}\right)
$$

où $x$ est le vecteur des variables explicatives du modèle, $q_2$ le nombre de neurones sur la couche cachée, $g$ la fonction de lien de la couche cachée (typiquement $g$ est la sigmoïde $g : x \rightarrow \frac{1}{1+e^{-x}}$), $w^{(1)}$  sont les poids entre la couche d'entrée et la couche cachée et $w^{(2)}$ les poids entre la couche cachée et la couche de sortie.
L'intérêt de ce type de réseau est expliqué par le résultat suivant (\cite{hornik_NN1993}) : les réseaux de neurones à une couche cachée permettent d'approcher, avec la précision souhaitée, n'importe quelle fonction continue (ou d'autres fonctions qui ne sont pas nécessairement continues) : c'est ce que l'on appelle la capacité d'approximateur universel et c'est aussi ce qui leur permet de s'appliquer avec une grande efficacité à un grand nombre de modèles.

\subsubsection{Modèle pour les données de Garrotxes}
Dans l'exemple des données de Garrotxes, plusieurs facteurs ont été retenus comme pouvant influencer l'occupation du sol d'un pixel donné à la date $t$ :
\begin{itemize}
\item[$\bullet$] \emph{pour l'aspect temporel (processus d'ordre 1)} : la valeur du pixel considéré à la date précédente (t - 1) que l'on exprime sous forme disjonctive (par exemple, si l'on dispose de 8 catégories d'occupation du sol, la première sera codée sous la forme d'un vecteur à 8 coordonnées : (1 0 0 0 0 0 0 0), la seconde : (0 1 0 0 0 0 0 0), etc) ;
\item[$\bullet$] \emph{pour l'aspect spatial} : la fréquence de chaque type d'occupation du sol dans le voisinage du pixel considéré à la date précédente ($t - 1$). Se pose alors le problème du choix du voisinage (taille et forme) : pour la forme, diverses possibilités s'offrent à nous, de la plus simple (voisinage carré ou en étoile) à des voisinages plus sophistiquées (voisinage suivant la pente pour mieux tenir compte des influences morphologiques du terrain). Quant à la taille du voisinage, il s'agira de déterminer jusqu'à quelle distance un pixel est susceptible d'influencer le pixel considéré. Afin de respecter la spatialisation de la carte, on pondèrera l'influence d'un pixel par une fonction décroissante de la distance au pixel considéré (cf. Figure \ref{geo2_voisinage}) ;

\begin{figure}[h]
\begin{center}
\includegraphics[width=10 cm]{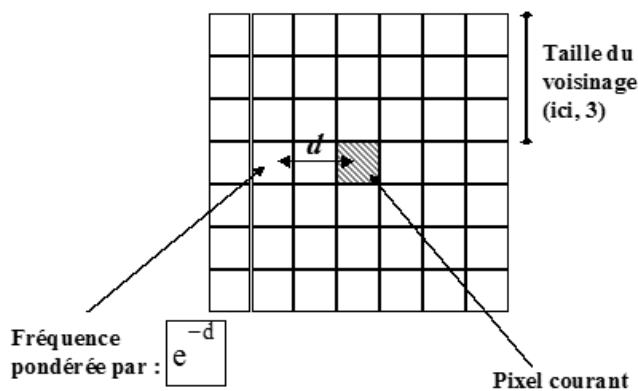}
\end{center}
\caption{\label{geo2_voisinage} Un exemple de voisinage}
\end{figure}

\item [$\bullet$] des variables environnementales (pente, altitude, \ldots).
\end{itemize}

\subsubsection{Mise en \oe uvre}
A l'issue d'une phase exploratoire qui nous a permis de cerner les variables pertinentes et divers paramètres du modèle comme la forme du voisinage (que, dans un soucis de simplicité, nous avons choisi carré), sa taille (finalement fixée à 3 ; cf. Figure \ref{geo2_nn}) ou le nombre de neurones optimal sur la couche cachée, l'architecture choisie compte, en entrée, 19 neurones :
\begin{itemize}
\item [$\bullet$] \emph{pour l'aspect temporel} : 7 neurones pour le codage disjonctif de la valeur du pixel (le bâti, constant, ayant été retiré de l'étude) à la date précédente $(t - 1)$ ;
\item [$\bullet$] \emph{pour l'aspect spatial} : 8 neurones pour la fréquence des divers types d'occupation du sol dans le voisinage (fréquence pondérée par une fonction décroissante de la distance) ;
\item [$\bullet$] 	4 neurones pour les variables d'environnement pente, altitude, exposition et distance aux infrastructures (préalablement centrées et réduites).
\end{itemize}

Le réseau dispose aussi de 8 neurones sur la couche cachée et de 7 neurones en sortie, chacun estimant la probabilité d'appartenance du pixel à un type d'occupation du sol (hors bâti).

Pour la phase d'apprentissage, nous avons utilisé comme jeu de données la carte de 1980 avec comme cible à prévoir la carte de 1989 et la carte de 1989 avec comme cible à prévoir la carte de 2000. Après avoir repéré de larges zones dans lesquelles l'occupation du sol était stable, nous avons considéré uniquement les pixels dont un voisin au moins avait une valeur d'occupation du sol différente (exploitant ainsi pleinement la spatialisation des données) : ces pixels seront, dans la suite, nommés pixels frontières. Les autres pixels ont été considérés comme stables dans un intervalle de temps de 10 ans.

D'un point de vue calculatoire, les programmes ont été réalisés à l'aide du logiciel Matlab (cf \cite{beale_demuth_NNTUG1998}) et sont disponibles sur demande.

\subsection{Approche par modèle linéaire généralisé}

Le \emph{modèle de régression logistique} est un modèle linéaire généralise (et donc paramétrique) dans lequel la variable réponse est qualitative et qui permet d'obtenir une prédiction de celle-ci en tenant compte d'un ensemble d'informations issues de variables explicatives. Lorsque la réponse possède plus de deux modalités, on parle de \emph{modèle de régression logistique multiple} ou \emph{modèle de régression polychotomique} (\cite{hosmer_lemeshow_ALR1989}). D'autres développements plus récents concernant ce modèle ont été réalises par \cite{kooperberg_bose_stone_JASA1997}. Ce type de modèle logistique est particulièrement bien adapte à notre problématique puisqu'il s'agit de prédire pour chaque pixel de la carte un type d'occupation du sol (variable réponse ayant 8 modalités codées par un entier $\nu$, $\nu = 1, 2, \ldots, 8$). La spécificité de notre étude vient du fait que, outre la nature topographique du pixel (pente, altitude, ...), le modèle choisi doit tenir compte d'un effet spatial (état de la végétation dans l'environnement du pixel) et d'un effet temporel (évolution du type d'occupation du sol du pixel et de son environnement). En ce sens il s'agit d'adapter le modèle de régression logistique à notre cadre, un des enjeux les plus importants étant le choix de la forme et de la taille de l'environnement pris en compte par le modèle.

De façon générale, le modèle de régression logistique permet de modéliser, en fonction d'un certain nombre de paramètres, la probabilité pour que le type d'occupation du sol d'un pixel au temps $t$ (c'est-à-dire la variable réponse) soit égal à un des 8 catégories d'occupation du sol. Il s'agit donc d'estimer les paramètres inconnus du modèle, et ensuite les probabilités a posteriori de type d'occupation du sol sachant les valeurs des différentes variables explicatives. On utilise ensuite une règle de type \emph{bayésien} consistant à affecter au temps $t$ à un pixel donné l'indice de végétation ayant la plus forte probabilité \emph{a posteriori}.

\subsubsection{Régression logistique multiple spatio-temporelle}

Indexons par $i = 1, 2, \ldots, N$ les pixels de la carte d'occupation du sol et notons  ${\cal I}_i$ l'ensemble des informations dont on dispose concernant le pixel numéro $i$. D'un point de vue formel, le modèle de régression logistique multiple que nous adoptons peut se présenter sous la forme générale suivante :
\begin{equation}
\label{geo2_model reglogis}
\log \left(\frac{Prob(\textrm{pixel}_i=\nu|{\cal I}_i)}{Prob(\textrm{pixel}_i=8|{\cal I}_i)}\right)=\alpha_\nu +\gamma_{\nu,{\cal I}_i},\ i=1,\ldots,N,
\end{equation}

où $\alpha_\nu$ est un paramètre associé au type d'occupation du sol $\nu$ que l'on souhaite prédire pour le $i^\textrm{ème}$ pixel et  $\gamma_{\nu,{\cal I}_i}$ un ensemble de paramètres lies à $\nu$ ainsi qu'aux informations concernant toujours ce pixel numéro $i$. Ainsi, le nombre total de paramètres mis en jeu dans ce modèle dépend uniquement du nombre de types d'occupation du sol et du nombre de variables explicatives. Dans l'expression (\ref{geo2_model reglogis}), $Prob(\textrm{pixel}_i=\nu|{\cal I}_i)$ représente la probabilité que l'occupation du sol du pixel $i$ soit du type $\nu$ lorsque les variables explicatives prennent les valeurs décrites par l'ensemble ${\cal I}_i$. Notons que l'expression (\ref{geo2_model reglogis}) modélise le rapport (son logarithme) de la probabilité qu'un pixel prenne la modalité v sur la probabilité que ce pixel prenne la modalité codée 8 ce qui permet d'intégrer la contrainte que la somme des huit probabilités est égale à 1.
Dans l'expression (\ref{geo2_model reglogis}), nous devons intégrer l'\emph{effet temporel} : celui-ci est pris en compte en faisant dépendre le type d'occupation du sol du pixel $i$ du temps $t$ c'est-à-dire en posant $\textrm{pixel}_i = \textrm{pixel}_i(t)$. Par ailleurs l'information (ou plus exactement une partie de cette information) dépend du temps $t - 1$ : ${\cal I}_i = {\cal I}_i(t - 1)$. L'idée consiste donc à calculer la probabilité qu'un pixel prenne un type d'occupation du sol $\nu$ à l'instant $t$ en fonction de l'information que l'on possède sur ce même pixel à l'instant précédent $ t - 1$ ; on répète cette procédure pour tous les pixels de la carte. Connaissant les cartes 1980 $(t - 1)$ et 1989 ($t$), on peut estimer l'ensemble des paramètres de notre modèle de sorte à ajuster au mieux la carte 1989. Il suffit alors d'incrémenter le temps dans notre modèle pour prédire la carte à l'instant futur $t + 1$  (2000) à partir de la carte observée à l'instant $t$ (1989).
Quant à l'\emph{effet spatial}, il est pris en compte de la même façon que dans l'approche par réseau de neurones. Il est en effet naturel de penser que l'évolution de l'occupation du sol du pixel $i$ dépend de celle des pixels environnants. Pour cela on considère un voisinage carré $V_i$ autour du pixel numéro $i$ que l'on souhaite prédire et on extrait comme information du voisinage $V_i$ le nombre de pixels prenant le type numéro 1 d'occupation du sol, le type numéro 2 d'occupation du sol, \ldots Cette façon de procéder revient à supposer une \emph{invariance isotrope}, c'est-à-dire que le type d'occupation du sol autour du pixel $i$ ne dépend pas de la direction. Dans la mise en oeuvre de la méthode, nous avons privilégié la simplicité de la forme du voisinage (carré). Lors de développements ultérieurs, on pourrait envisager d'autres formes que le carré (étoile, rectangle, ...) et varier s'il en résulte un gain ou pas en terme de prédiction. On peut également envisager une modélisation privilégiant certaines directions c'est-à-dire rompant avec l'hypothèse d'invariance isotrope. Notons cependant qu'il en résulterait un modèle avec un plus grand nombre de paramètres et que de ce point de vue on doit également composer avec la capacité à bien estimer un modèle qui serait trop complexe.

En combinant effet temporel et effet spatial, on est finalement amené à considérer le modèle suivant
$$
\log \left(\frac{Prob(\textrm{pixel}_i(t)=\nu|{\cal I}_i(t-1))}{Prob(\textrm{pixel}_i(t)=8|{\cal I}_i(t-1))}\right)=\alpha_\nu +\gamma_{\nu,{\cal I}_i(t-1)},\
$$
où ${\cal I}_i(t - 1)$ englobe l'information extraite du voisinage $V_i(t - 1)$, c'est-à-dire tient compte de l'occupation du sol observée autour du pixel numéro $i$ à l'instant $t - 1$.
Enfin, ${\cal I}_i(t - 1)$ comprend également l'information issue des variables telles que la pente ou l'altitude décrites plus haut.

\subsubsection{Mise en oeuvre}

D'un point de vue pratique, la mise en oeuvre se décompose en deux étapes : une étape d'estimation et une étape de validation.

\emph{Etape d'estimation} : On estime les paramètres du modèle ($\alpha_\nu$ et ceux contenus dans $\gamma_{\nu,{\cal I}_i(t-1)}$). La procédure d'estimation est basée sur la maximisation de la \emph{vraisemblance pénalisée}, critère bien connu en statistique pour la stabilité des solutions obtenues. L'algorithme d'optimisation utilise est de type \emph{Newton-Raphson}. Remarquons que la pénalisation introduit un nouveau paramètre, appelle paramètre de pénalisation et note $\epsilon$, qu'il faudra choisir. Comme cela a été dit précédemment, on utilise les cartes 1980 et 1989 pour estimer les paramètres, ceci pour différentes tailles de voisinage et valeurs $\epsilon$.

\emph{Etape de validation} : Il s'agit de déterminer la taille de voisinage et le paramètre de pénalisation optimaux en ce sens que ces choix fourniront une prédiction de la carte 2000 la plus proche possible de celle observée Plus précisément, à l'aide de étape précédente, il est possible de construire plusieurs prédictions de la carte 2000, chacune correspondant à différentes tailles de voisinage et valeurs $\epsilon$. En comparant les cartes prédîtes pour 2000 avec la carte réelle de 2000, on repère la carte qui possède le plus petit nombre de pixels mal prédits; la taille de voisinage et le paramètre de pénalisation correspondants seront considérés comme étant optimaux.

\section{Résultats et interprétation}

Les trois approches ont été testées par la modélisation de l'occupation du sol à la dernière date connue (2000).

\begin{figure}[h]
\makebox[0.2 cm]{}
\makebox[6 cm][r]{\includegraphics[width=6 cm]{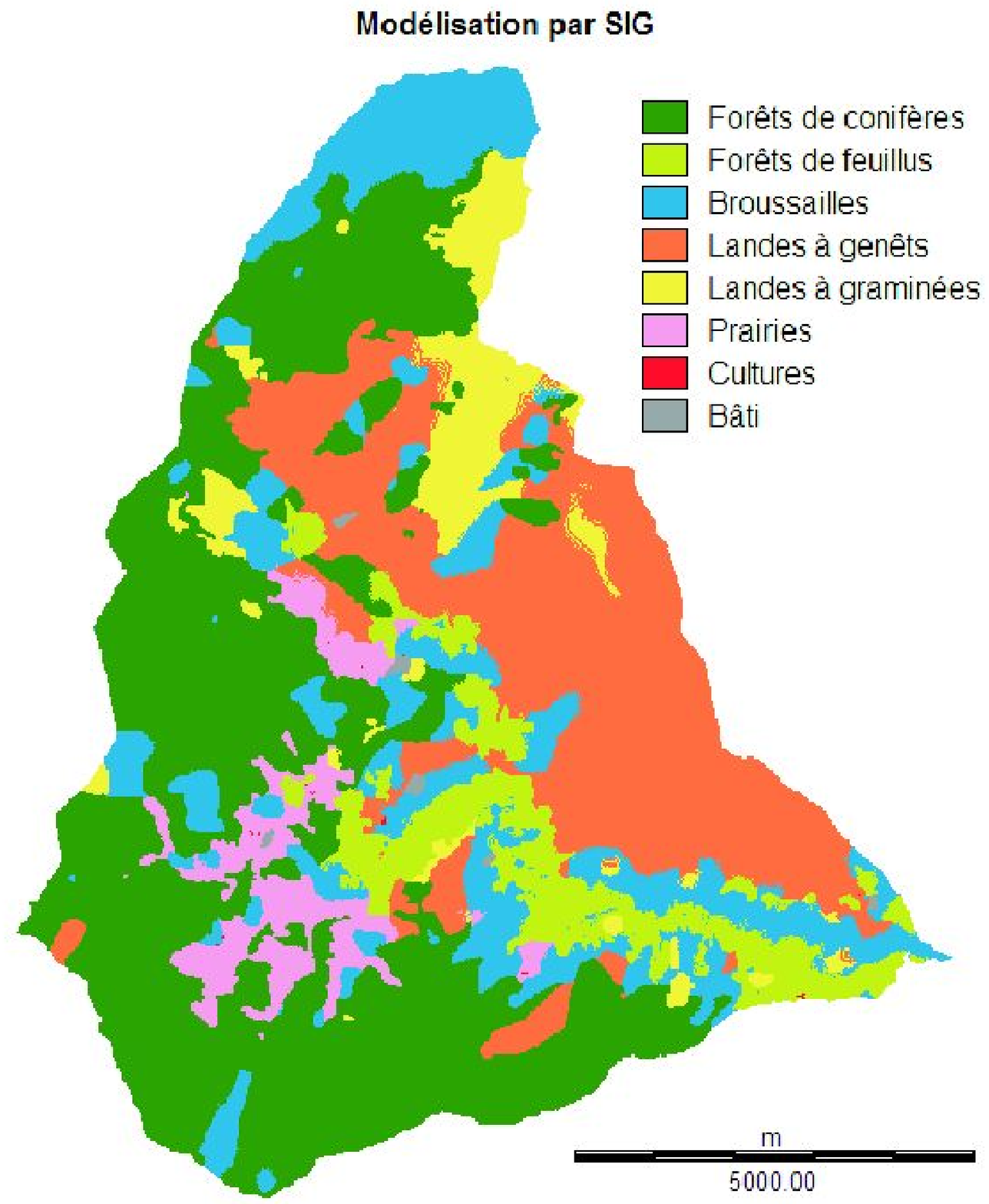}}
\makebox[0.2 cm]{ }
\makebox[6 cm][l]{\includegraphics[width=6 cm]{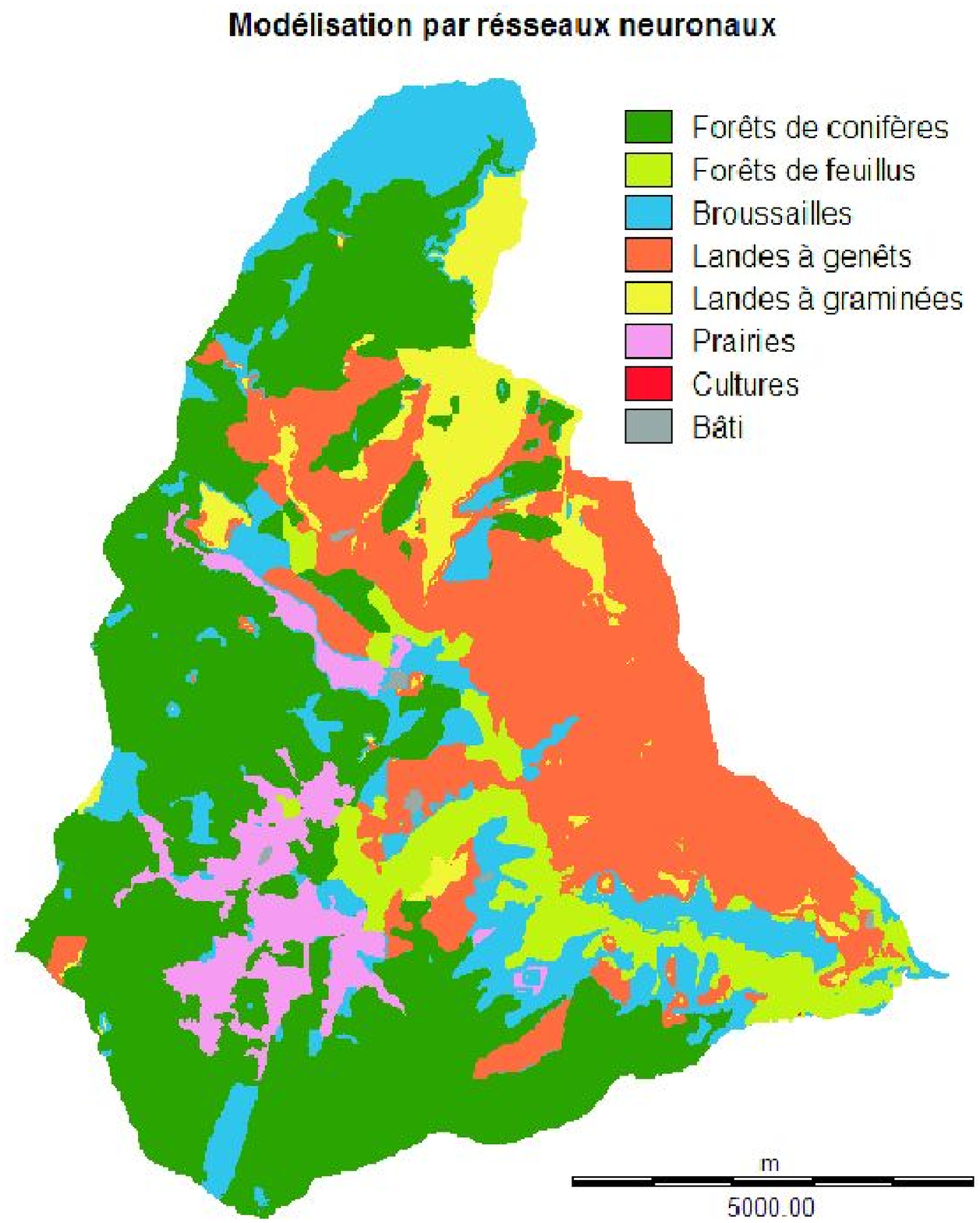}}\\
\makebox[4 cm]{}\\
\makebox[0.2 cm]{}
\makebox[6 cm][r]{\includegraphics[width=6 cm]{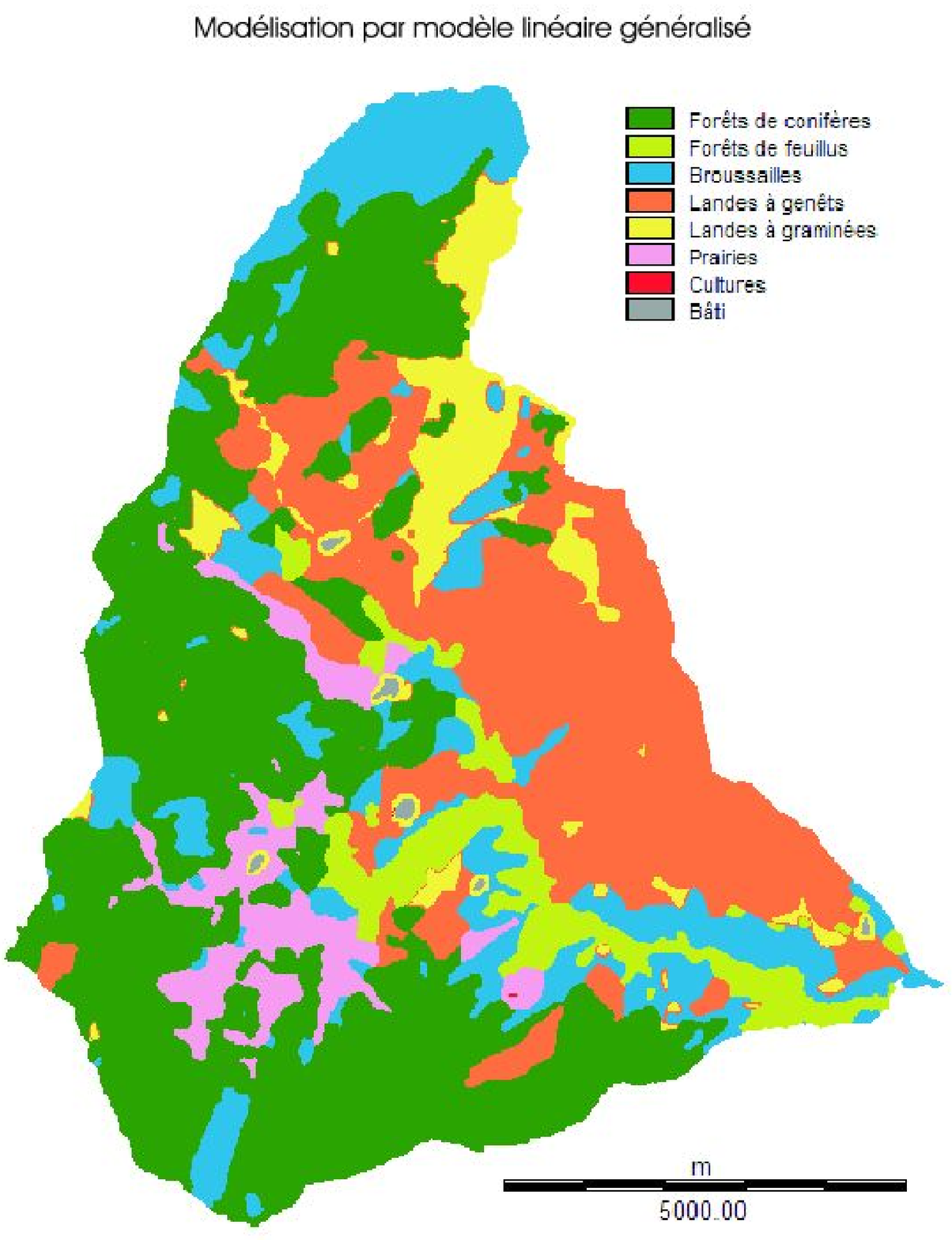}}
\makebox[0.2 cm]{ }
\makebox[6 cm][r]{\includegraphics[width=6 cm]{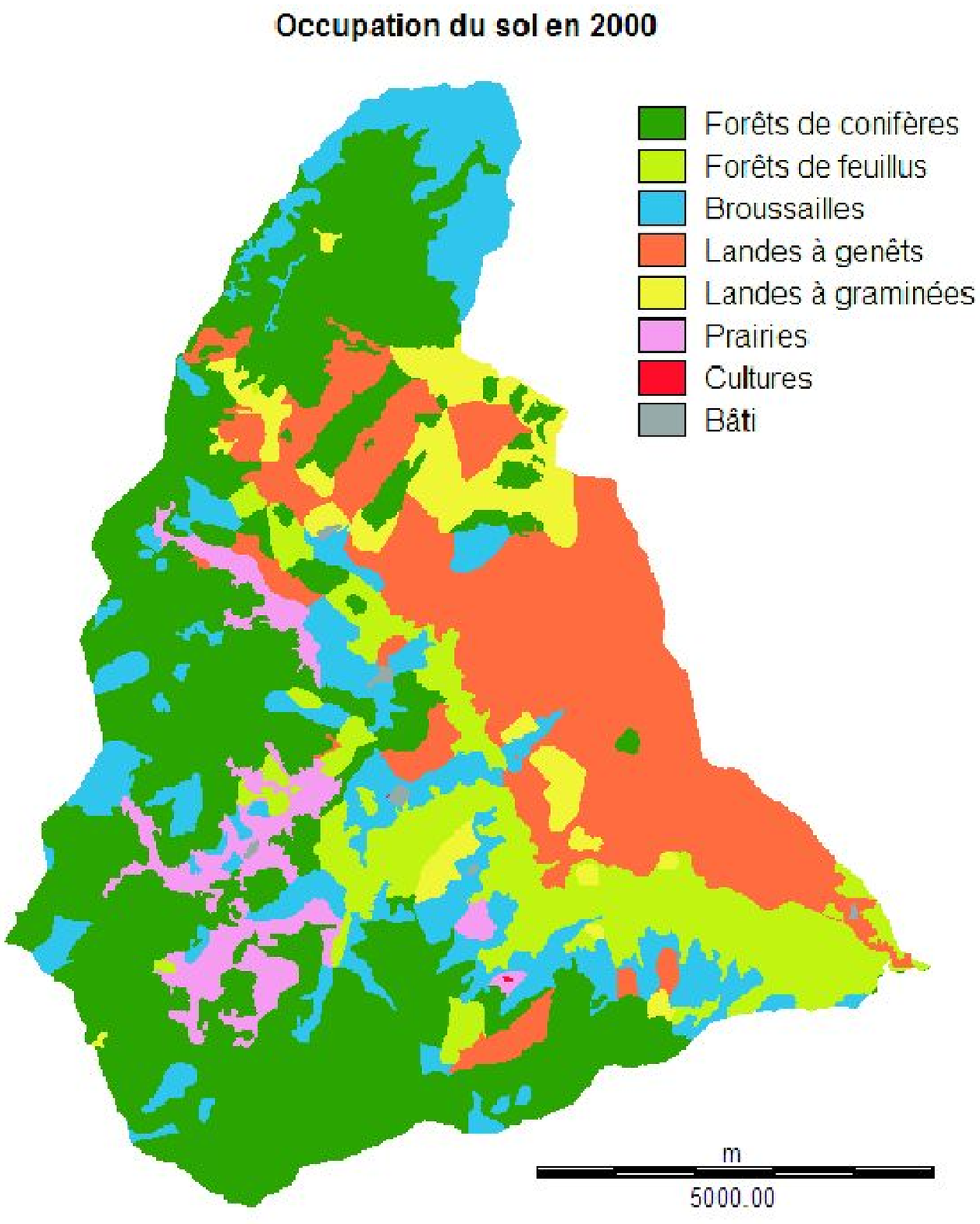}}
\caption{Résultats des modélisations de l'occupation du sol en 2000 et occupation du sol réelle}
\label{geo2_prev}
\end{figure}

Les résultats globaux (pourcentage de surface) de ce test sont proches de la réalité (cf. Figure \ref{geo2_prev} et Tableau \ref{geo2_tab surface}).

\begin{table}[h]
\begin{tabular}{|c|c|c|c|c|}
\hline
Occupation du sol en 2000 & Réalité & \multicolumn{3}{|c|}{Modélisation}\\
& & SIG & Réseaux neuronaux & Modèle linéaire généralisé\\
\hline
Forêt de conifères & 40,9 & 40,7 & 41,4 & 41,2\\
\hline
Forêt de feuillus & 11,7 & 6,0 & 7,2 & 6,3\\
\hline
Broussailles & 15,1 & 16,9 & 14,1 & 14,6\\
\hline
Landes à genêts & 21,6 & 23,3 & 25,1 & 25,9\\
\hline
Landes à graminées & 5,7 & 7,6 & 6,0 & 6,2\\
\hline
Prairies & 4,8 & 5,2 & 6,0 & 5,7\\
\hline
Cultures & 0,01 & 0,1 & 0 & 0\\
\hline
\end{tabular}
\caption{\label{geo2_tab surface} Surface en pourcentage de l'occupation du sol en 2000, réelle et modélisée}
\end{table}

Cependant il convient d'analyser la répartition spatiale de ces sommes de surface prédites par catégorie à haute résolution (coté du pixel environ 18 m). Le Tableau \ref{geo2_tab pourcentage} compare les résidus par catégorie en pourcentage de la surface réelle de chaque catégorie mise à part les cultures dont le nombre de pixels tend vers zéro. Le premier enseignement de ce tableau est la relative concordance des résultats des trois approches. On remarque également que la modélisation de modalités ayant une grande superficie (forêt de conifères, landes à genêt) est plus facile que celle des catégories d'occupation du sol de faible ampleur. Ainsi, quel que soit le modèle, moins d'un pixel sur deux a été correctement prédit pour la catégorie broussailles. Parmi les modalités de faible surface, on note cependant de grandes différences selon leur stabilité spatio-temporelle : les prairies étant plus stables que les landes à graminées, leur taux de prédiction est meilleur. 

Les taux de prédiction globale des trois méthodes sont très proches : 72.8 \% (SIG), 74.3~\% (réseaux neuronaux) et 72.8 \% (modèle linéaire généralisé).

\begin{table}[h]
\begin{tabular}{|c|c|c|c|c|c|c|}
\hline
Occupation & Forêt de & Forêt de & Brous- & Landes & Landes à & Prairies\\
du sol en 2000 & conifères & feuillus & -sailles & à genêts & graminées & \\
Surface (\%) & 40,9 & 11,7 & 15,1& 21,6 & 5,7 & 4,8\\
\hline
Résidus (\%) & & & & & & \\
de modélisation & & & & & & \\
SIG (27,2\%) & 11,42 & 55,28 & 51,92 & 17,13 & 54,39 & 30,35\\
\hline
Réseaux de & 10,60 & 45,84 & 54,54 & 16,23 & 59,38 & 19,26\\
neurones (25,7\%) & & & & & & \\
\hline
Modèle linéaire & 11,88 & 51,65 & 57,07 & 14,35 & 59,24 & 25,57\\
généralisé (27,2 \%) & & & & & & \\
\hline
\end{tabular}
\caption{\label{geo2_tab pourcentage} Pourcentage de résidus par catégorie d'occupation du sol et approche modélisatrice}
\end{table}

Les modélisations n'ont pas vocation à prédire la réalité mais peuvent nous aider à mieux comprendre des changements spatio-temporels environnementaux et sociaux complexes. Dans ce sens, l'interprétation des résultats des modélisations doit tenir compte des limites des modèles. La modélisation de l'occupation du sol signifie une simulation de ce que la réalité pourrait être, un scénario raisonné et quantifiable dans le contexte d'aide à la décision.

Cependant une interprétation minutieuse des résultats devrait nous permettre à améliorer le modèle et, par conséquent, le taux de prédiction. Dans ce sens, l'analyse focalise surtout sur les résidus.

La catégorie d'occupation du sol la plus représentée (les conifères) obtient un très bon score de prédiction alors que les broussailles, relativement présentes sur le territoire, obtiennent un très mauvais score (plus de la moitié de mal prédits). Diverses remarques permettent d'expliquer ces phénomènes et de penser à des stratégies d'amélioration de la prédiction. Tout d'abord, on peut constater que les broussailles sont la catégorie naturellement la plus dynamique sur un territoire caractérisé par un équilibre entre espace forestier et pastoral régi notamment par la gestion pastorale. Les broussailles sont également soumises le plus à des effets aléatoires : un feu de forêt, une coupe ou bien l'abandon de pâturage transforment en l'espace de 10 ans une parcelle en broussailles ; ce sont des phénomènes complètement incontrôlables.

\begin{table}[h]
\begin{tabular}{|c|c|c|c|}
\hline
Ecart de prédication &\makebox[2 cm]{SIG} & Réseaux neuronaux & Modèle linéaire généralisé \\
\hline
1 catégorie & 12,9 & 12,5 & 13,0\\
\hline
2 catégories & 9,1 & 8,5 & 9,2\\
\hline
3 catégories & 3,2 & 2,9 & 3,1\\
\hline
4 ou 5 catégories & 1,9 & 1,8 & 1,9\\
\hline
Total résidus & 27,2 & 25,2 & 27,2\\
\hline
\end{tabular}
\caption{\label{geo2_tab residus} Analyse des résidus de la modélisation par l'écart de catégorie entre la réalité et la modélisation (données en pourcentage de la surface totale)}
\end{table}

Bien que l'occupation du sol soit décrite de manière qualitative, ses différentes catégories s'échelonnent entre des formations fermées (forêt de conifères, forêt de feuillus) et ouvertes (cultures). Ces rangs « paysagers » permettent de quantifier l'erreur de prédiction, exprimée en nombre de catégories (cf. Tableau \ref{geo2_tab residus}). Ainsi, quelle que soit la méthode de modélisation, pour environ la moitié des pixels mal prédits, l'erreur de prédiction n'est que d'une catégorie à une résolution spatiale élevée. Le nombre de résidus décroît fortement avec l'augmentation de l'écart entre la réalité et la projection.

\begin{table}[h]
\begin{tabular}{|c|c|c|c|c|c|c|c|c|}
\hline
& 3 modèles & \multicolumn{3}{|c|}{2 modèles} &\multicolumn{3}{|c|}{1 modèle} & Aucun\\
&&&&&&&& modèle\\
\hline

Prédiction & & RN & SIG & SIG & SIG & RN & MLG & \\
correcte par : &&+ MLG &+ MLG &+ RN &&&&\\
\hline
Forêt & 85,35 & 1,65 & 0,40 & 0,68 & 0,96 & 1,75 & 0,76 & 8,53\\
de conifères &&&&&&&&\\
\hline
Forêt & 46,26 & 0,90 & 0,57 & 3,11 & 4,98 & 3,93 & 0,50 & 39,75\\
de feuillus &&&&&&&&\\
\hline
Brous-& 32,38 & 5,92 & 2,64 & 2,50 & 7,75 & 3,89 & 1,03 & 43,89\\
-sailles &&&&&&&&\\
\hline
Lande & 76,98 & 4,99 & 2,45 & 0,74 & 2,82 & 0,70 & 0,93 & 10,39\\ 
à genêts &&&&&&&&\\
\hline
Lande& 26,30 & 7,63 & 3,27 & 2,24 & 7,71 & 2,91 & 2,32& 47,62\\
à graminées &&&&&&&&\\
\hline
Prairies & 59,14 & 11,64 & 1,66 & 4,70 & 1,06 & 5,26 & 1,99 & 14,55\\
\hline
\textbf{Total} & \textbf{66,49} & \textbf{3,75} & \textbf{1,42} & \textbf{1,54} & \textbf{3,23} & \textbf{2,33} & \textbf{0,92} & \textbf{20,32}\\
\hline
\end{tabular}
\caption{\label{geo2_tab relation}Mise en relation des scores de prédiction correctes des trois modèles avec l'occupation du sol en 2000. Données en \% de la surface totale. RN = modèle par réseaux neuronaux ; MLG = modèle linéaire généralisé ; SIG = modèle SIG }
\end{table}

\begin{figure}[h]
\begin{center}
\includegraphics[width=10 cm]{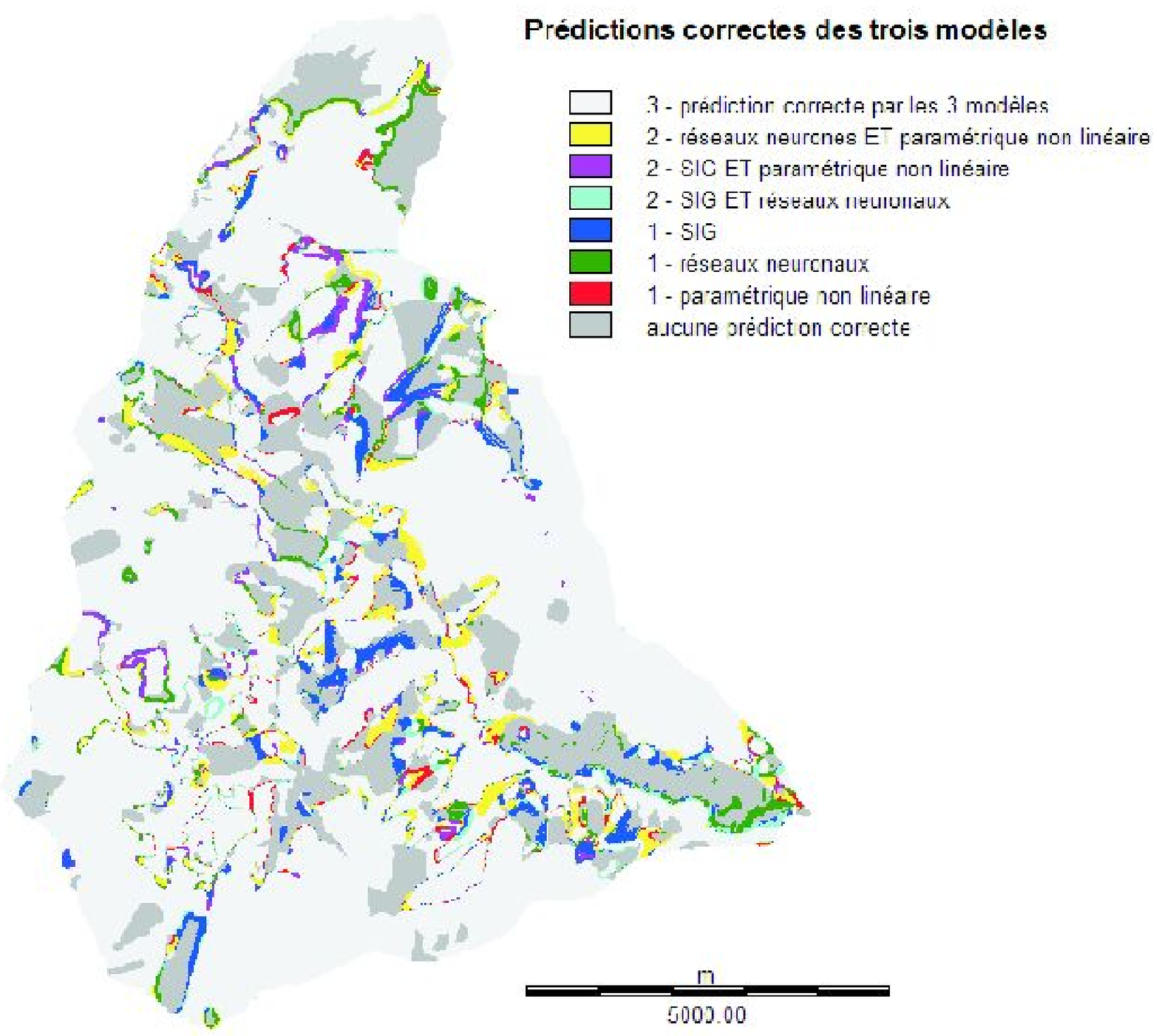}
\end{center}
\caption{\label{geo2_carte croisee} Prédictions correctes croisées par les 3 modèles}
\end{figure}

Un autre aspect intéressant est la grande concordance des trois modèles (cf. Figure \ref{geo2_carte croisee} et Tableau \ref{geo2_tab relation}). Ainsi 66.5 \% de la surface totale est correctement prédite par chacun des modèles, 20.3 \% par aucune.

Un autre enseignement du croisement des pixels correctement prédits par les différents modèles est la similitude des résultats des deux modèles statistiques. Globalement, les surfaces correctement prédites par deux modèles sur trois le sont le plus souvent par les réseaux neuronaux et le modèle linéaire généralisé (3.75 \%). Le taux de prédiction combiné de la méthode SIG avec chacun des modèles statistiques est nettement plus faible. La relative distinction du modèle SIG est corroborée par les surfaces correctement prédites par seulement une des approches (3.23 \%). Une analyse catégorielle souligne ce constat. Ainsi, pour les surfaces correctement prédites par uniquement deux modèles, cinq catégories d'occupation du sol sur six le sont par les approches statistiques. A l'inverse, les surfaces uniquement prédites correctement par un seul modèle le sont le plus souvent par le modèle SIG (quatre catégories d'occupation du sol sur six). En outre, le modèle SIG se distingue par le fait que son taux de prédiction est meilleur pour des zones affectées par un changement d'état, soit entre la dernière date de la phase d'apprentissage et la date simulée, soit durant la phase d'apprentissage. Les deux méthodes statistiques, au contraire, aboutissent à des résultats légèrement plus proches de l'observation sur les secteurs stables. Ce comportement spécifique sur les marges, s'explique par les procédures d'affectation spatiale différentes des transitions temporelles où le modèle SIG fait appel à une analyse géographique de la rugosité de l'espace par évaluation multicritère. Il s'agit, en l'occurrence, d'indices à approfondir dans la perspective de l'intégration des trois modèles en un seul.

Les résultats font également ressortir certaines limites des modélisations :
\begin{itemize}
\item[$\bullet$] La variable modélisée est décrite par un nombre limité de catégories. Ce choix, intentionnel, est motivé par la qualité et l'origine des données source afin de minimisez les erreurs d'interprétation. L'inconvénient est une certaine variabilité à l'intérieur de chaque catégorie qui n'est pas prise en compte par les modèles. Ainsi peut-on voir dans les trois cartes de modélisation (Figure \ref{geo2_carte croisee}) une vaste zone prédite en broussailles dans le secteur sud-est qui est réellement occupée par des bois de feuillus. Pendant la période d'apprentissage cette zone a été photointerprétée comme broussailles en 1980 et en 1989. Cependant les broussailles se sont densifiées, leur composition floristique a changé au profit de Quercus ilex formant une strate arborescente dominante en 2000 où la zone a été classée forêt de feuillus.
\item[$\bullet$] La période d'apprentissage est peu fournie. Les modélisations ne se basent que sur deux dates connues. Ceci pose des problèmes par rapport à la source de données à utiliser. En l'occurrence il s'agit de données haute résolution : des missions de photographies aériennes espacées dans le temps au point que l'utilisation de missions plus anciennes nous semble problématique sachant que le contexte socio-économique a beaucoup évolué.
\item[$\bullet$] Le taux d'explication de la variabilité spatio-temporelle de l'occupation du sol par les variables d'environnement est inégal selon les catégories d'occupation du sol.
\item[$\bullet$] Finalement chaque modèle est affecté par un bruit aléatoire. Des effets aléatoires comme des incendies de forêt, du chablis ou des programmes de reforestation sont difficilement prévisibles.
\end{itemize}

A cela s'ajoutent des limites spécifiques à chacun des modèles. Pour le modèle SIG deux facteurs limitants sont à mentionner.  Des zones stables pendant la période d'apprentissage sont prédites stables par l'analyse des chaînes de Markov. En outre,  les procédures EMC, EMO et l'automate cellulaire ne gèrent que ra répartition spatiale des scores de probabilités calculés par l'ACM. Cette restriction est, en partie, aussi valable pour les approches statistiques.

\section{Perspectives}

Les résultats exposés sont les premiers au sein d'un projet de recherche portant sur trois sites d'études. Les mêmes modèles seront appliqués à la Montagne de Lure en Haute Provence ainsi qu'à la Alta Alpujarra Granadina formant la partie occidentale du versant sud de la Sierra Nevada (Espagne).

Cette comparaison croisée entre plusieurs approches appliquées sur plusieurs sites devrait, d'une part, limiter l'effet de singularité propre à chaque terrain d'études et, d'autre part, nous donner à terme des indications afin de proposer un modèle intégré, issu des trois méthodes mises en \oe uvre actuellement parallèlement.
En même temps, les modèles restent évolutifs au sens où l'interprétation des premiers résultats réoriente les prochaines étapes. A ce propos il nous semble intéressant de remédier au problème de la variabilité interne à chaque catégorie d'occupation du sol par une approche semi quantitative. Aux modalités qualitatives décrites s'ajoutent des données ordonnées sur le taux de recouvrement de la strate arborescente.

Enfin, les approches mises en \oe uvre sans intervention « spécialiste » (du type SIG) donnent des résultats aussi bons que la méthode « supervisée ». La combinaison de méthodes purement mathématiques avec un guidage expert pourrait permettre de gommer les imperfections du modèle. Cette voie est encore à explorer afin de progresser vers un modèle intégré de la simulation prospective de l'occupation du sol.

\section{Bibliographie}
\bibliographystyle{alpha}
\bibliography{bibli-these}

\end{document}